# WIRELESS PUBLIC KEY INFRASTRUCTURE FOR MOBILE PHONES

Balachandra Muniyal[1]    Krishna Prakash[2]    Shashank Sharma[3]

[1]Dept. of Information and Communication Technology,
Manipal Institute of Technology, Manipal University, Manipal, India.
`bala_muniyal@yahoo.com`

[2]Dept. of Information and Communication Technology,
Manipal Institute of Technology, Manipal University, Manipal, India.
`kkp_prakash@yahoo.com`

[3]Dept. of Information and Communication Technology,
Manipal Institute of Technology, Manipal University, Manipal,India.
`shanks406@gmail.com`

*ABSTRACT- Mobile devices play an important role in the wireless network environment for providing different services over internet. The business transactions over wireless electronic devices are not secure and hence the messages are prone to be intercepted and modified by an intruder. So, devices supporting wireless internet must be guaranteed at the same level of security as the wired network. PKI (Public Key Infrastructure) used in the wired environment is not suitable for wireless environment because of the less powerful processor and small memory. This arises a need for the development of a Wireless Public Key Infrastructure (WPKI) that provides the similar security level as the wired PKI suitable for mobile phone. In this paper, a discussion of public key infrastructure and an experimental set up for Wireless Public key Infrastructure for mobile phones are made.*

*KEYWORDS*—WPKI,    Digital Certificate,    Wireless internet,    M-commerce

## 1. INTRODUCTION

Several kinds of e-services are provided by mobile phones over internet. Due to unreliable wireless media, dynamic topology and limitations on infrastructure these services are vulnerable to security threats and also impose restrictions on the usage of conventional PKI. PKI is a security architecture that provides high level of confidence for the information to be exchanged over internet. To guarantee security of Mobile commerce(M-commerce) via wireless internet, public key infrastructure technology suitable for wireless environment must be required. Wireless internet has many restrictions compared to wired one. A mobile phone does not have the same computational ability and storage capacity as a desktop computer, and wireless communication has lower transmission bandwidth than its wired counterpart [1-4]. Applying wired internet protocols to mobile phone has many problems such as the limitations in screen size, computing power, memory capacity and these limitations should be addressed while designing PKI environment for the wireless devices. Optimization of certification management protocols and size of the data processed in the mobile device are the major factors to be considered in the design which has a considerable reduction in the size of the module to be installed in the mobile phone.

## 2. PUBLIC KEY INFRASTRUCTURE

A Public Key Infrastructure (PKI) is a system consisting of set of hardware and software used for the management of public key and distribution of digital certificates which are used to verify a particular public key belongs to a certain entity. The PKI creates digital certificates which map public keys to entities, securely stores these certificates in a central repository, and revokes them if needed when it is not in use.

Public key cryptography is used to transmit user's public key in PKI environment. Public key of the user is advertised and corresponding private key kept secret.

A PKI consists of Certification Authorities (CAs), Registration Authorities (RAs), Certificate holders, Clients, Repositories, Cryptographic Algorithms and Protocols. As shown in Figure 1, validating the identity of user and then issuing a digital certificate is done by CA. RA is a third party that validates the identity of those applying for a digital certificate. The PKI directory contains information regarding the CA's digital certificate[2][3].

A Public Key Infrastructure ensures the following:

- Ensures the quality of information transmitted over the network.
- Lifetime and validity of the information.
- Certainty of the privacy, and source and destination of that information of that information.
- To ensure non repudiation .

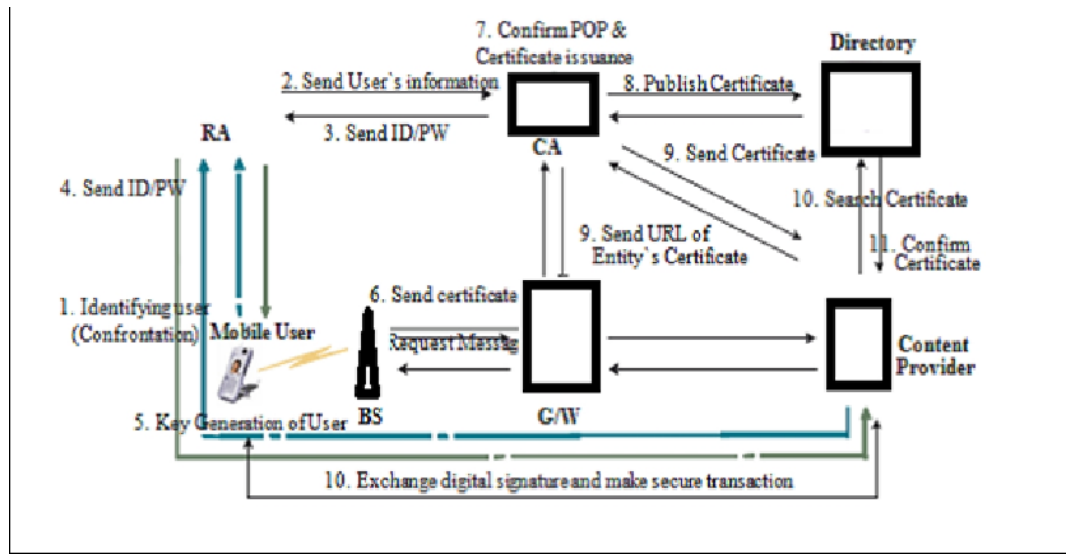

Figure 1 - PKI Model and Scenario

.

## 3. IMPORTANCE OF WIRELESS PUBLIC KEY INFRASTRUCTURE

Wireless devices have some technical limitations such as less powerful CPU, less memory size, restricted battery power, small display and input devices compared to devices in wired infrastructure. A mobile phone lacks computing capabilities of PKI services such as key generation, digital signature generation and verification, certificate validation, Certificate Revocation List (CRL) verification, memory size of storing certificate and CRL. Because of the technical limitations of wireless environment, processing of CMP (Certificate Management

Protocol) in the mobile phone, and downloading CRL required for certificate verification becomes very hard.

Following paragraphs discuss how wireless public key infrastructure can be deployed for mobile phones.

## 3.1 Wireless PKI for mobile phones

WPKI through wireless internet with the same level of security as that of wired internet, the following requirements must be satisfied:

- Select optimal digital signature algorithm to be calculated in mobile phone.
- Small piece of data to be stored in mobile phone and transmitted over networks.
- Optimize CMP protocol to be processed in mobile phone and through wireless bandwidth.
- Optimize certificate validation scheme.

### *3.1.1 Optimal Digital Signature*

Following are the two important algorithms for the generation of optimal signature:

- ECDSA (Public-Key Algorithm)
- DES (Symmetric-Key Algorithm)

The principal attraction of Elliptic Curve Digital Signature Algorithm is that, it offers equal security for a small key size as that of RSA, thereby reducing processing overhead. ECC based Elliptic Curve Digital Signature Algorithm generates 163-bit key size, equivalent to RSA 1024-bit key size, it takes shorter time to generate public key pair in mobile phone than RSA algorithm. Since ECDSA 163-bit key size is less than RSA 1024-bit key size, a certificate size including the public key could be reduced [5].

The Data Encryption Standard (DES) played a vital role in the area of cryptography for the encryption of electronic data . It is no longer secure in its original form, but in a modified form it is still useful. DES is now considered to be insecure for many applications. This is chiefly due to the 56-bit key size being too small [6].

### *3.1.2 Optimal Certificate Profiles*

We describe Wireless X.509 certificate profile issuing for mobile phone and server, and short-lived certificate profile issuing for server to reduce verification load of mobile phone.(Figure 2 and Figure3)[3].

X.509 certificate consists of basic field and extension field. Generation implies that a certificate has to include the specified field, and process implies that if the specified filed is present in the certificate, the field must be examined when the certificate is verified. Basic purpose is to reduce the size of the certificate by eliminating unnecessary and unused fields from the certificate.

Wireless X.509 certificate profile for mobile phone

|  | Generation | Process |
|---|---|---|
| *Basic field* | | |
| Version | m | m |
| Serial number | m | m |
| Signature | m | m |
| Issuer | m | m |
| Validity | m | m |
| Subject | m | m |
| Subject Public Key Info | m | m |
| Issuer unique identifier | x | x |
| Subject unique identifier | x | x |
| *Extension field* | | |
| Authority key identifier | m | o |
| Subject key identifier | m | o |
| Key usage | m | m |
| Private key usage period | x | x |
| Certificate policy | m | m |
| Policy Mapping | – | – |
| Subject alternative names | m | m |
| Issuer alternative names | o | m |
| Subject directory attributes | x | x |
| Basic constraints | x | x |
| Name constraints | – | – |
| Policy constraints | – | – |
| Extended Key Usage | o | m |
| CRL distribution points | m | o |
| Domain information | o | o |
| Authority information access | m | o |

m: mandatory, o: optional, x: not recommended, –: not defined.

**Figure 2 - Wireless X.509 Certificate**

Short-lived certificate profile

| Field name | Value | Generation/process |
|---|---|---|
| certificate_version | V1 | m |
| signature_algorithm | ECDSA with SHA | m |
| issuer | ⟨Text⟩ | m |
| valid_not_before | GMT | m |
| valid_not_after | GMT | m |
| Subject | ⟨Text⟩ | m |
| public_key_type | ECDH | m |
| parameter_specifier | optaion | m |
| signature | ECDSA signature value | m |

GMT: Greenwich Mean Time, ECDH : EC-based Diffie-Hellman.

**Figure3 - Short-lived Certificate**

### 3.1.3 Optimal Certificate Request and Management Protocol

The mobile phone has to send a secured request for the certificate to a CA and CA issues it to the mobile phone. The followings are requirements of certificate request protocol:
- Certificate request message is constructed at mobile phone. This value should include a public key, end-entity's reference number like ID and password.
- Corresponding to the public key for which a certificate is being requested, a Proof of Possession of the private key value is included in certificate request message
- Method that the certificate request message is securely communicated to a CA.

To satisfy these requirements, wireless certificate management protocol is developed (Figure4).

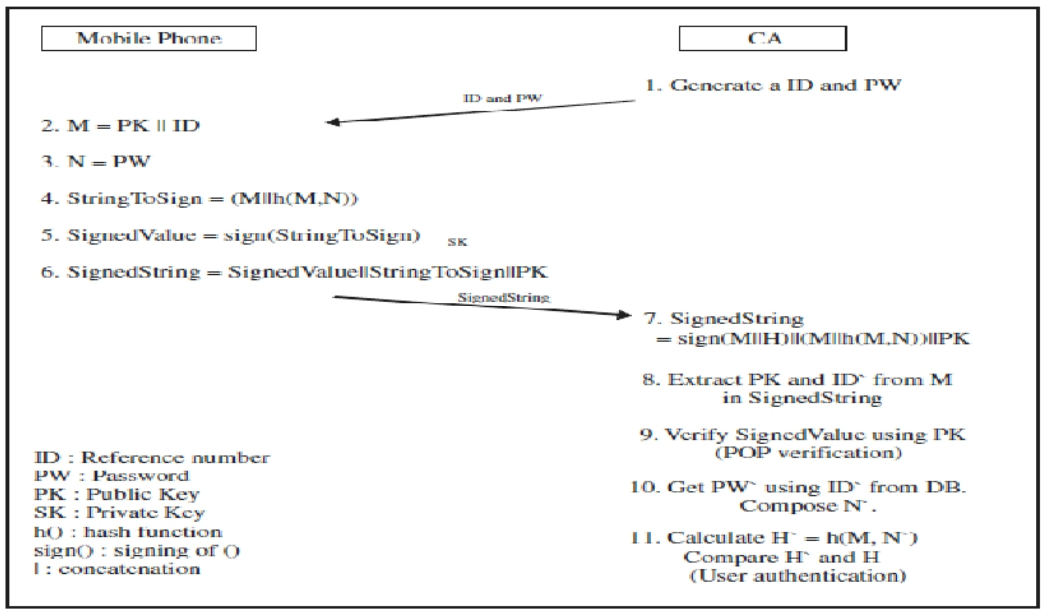

Figure. 4 - Certificate Management Protocol.

### *3.1.4. Optimal Validation Scheme*

A mobile phone delegates validation authority to a third party server such as OCSP (Online Certificate Status Protocol) to validate certificate in this model. As shown in Figure 6, the mobile phone can avoid burden of CRL download and storage as well as the complicated procedure to acquire and verify certificate chain.

## 4. Sequence of operations in the proposed set-up

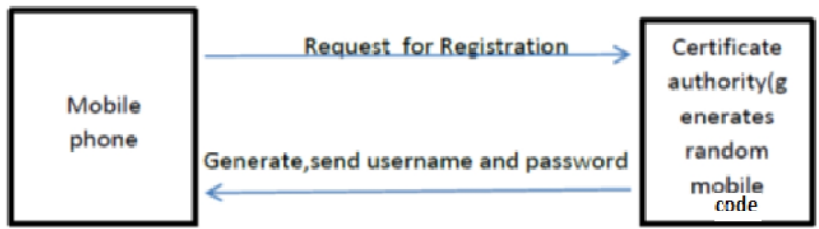

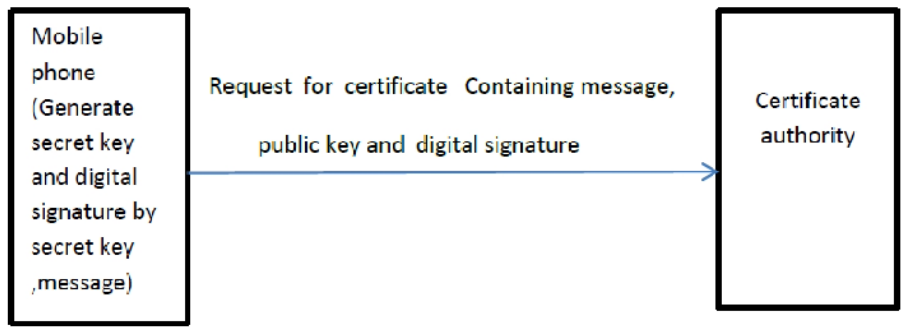

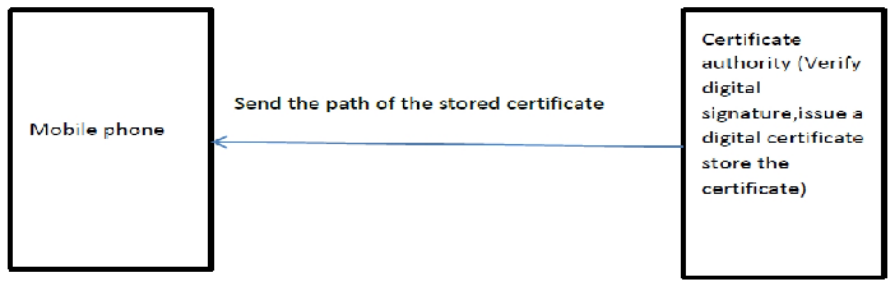

Figure 5- Activities between CA and Mobile phone for generating certificate.

New mobile user (Client) requests for registration at certification authority. A request for registration is received at CA (Server) and a random code is generated specific for each requesting wireless device. The CA then generates a username and password specific to the user and stores the values along with the random code in the database. The user supplied with username and password. The user can now generate a public key-private key/secret key using a digital signature algorithm. A message is then entered by the user or the username and password and public/secret key is used as the message. The message digest is generated using a hash algorithm. Using the message digest and the private/secret key the digital signature is generated. The digital signature along with the message and public/secret key is sent from the user to the CA for verification purposes. The digital signature is then unlocked using the public/secret key of the user at CA and the message digest is matched with the message digest generated on the CA. Once the digital signature is verified, CA issues a Digital Certificate for the user. The values are stored in a notepad file and this file is stored in a directory assigned to CA. The URL/Pathname of the certificate in the directory is sent to the user for future transactions with other servers and thus reducing the load on wireless devices such as mobile phones. Whenever a transaction happens between mobile phone and a server, initially, the server sends its certificate to the mobile. The mobile sends this certificate to the OCSP server. OCSP server then downloads the CRL list from the directory and checks for certificate validation. Once validated, the OCSP server sends its response to the mobile which then carries on the future transactions with the mobile device thus reducing future transactions. As shown in Figure 6, the server on receiving the Certificate URL of the mobile, sends it to OCSP server for validation. In turn, the CRL is sent to the CA.

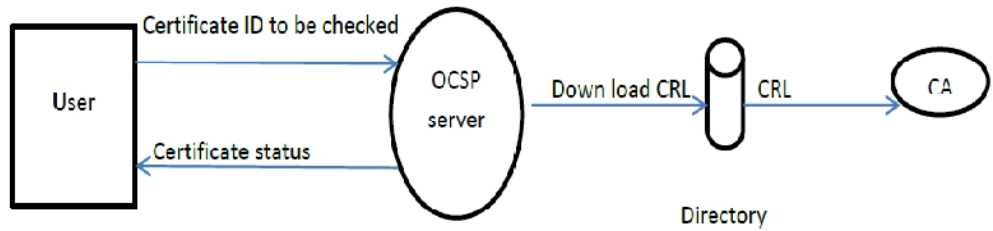

Figure 6: Interaction with OSCP server

## 5. CONCLUSIONS

In this article, the concept of PKI is discussed briefly and the limitations of PKI for wired network is addressed. For the wireless environment, same security level as wired PKI supporting mobile phone is considered. The proposed wireless PKI model aimed at secure M-commerce based on mobile phone through wireless communication.

To reduce the complexity of certificate validation, the optimal certificate profile for X.509 and short-lived certificate, which reduced the size of the certificate, and applied OCSP model to efficiently validate X.509 certificate in mobile phone.

WPKI services within the WAP environment, must be optimized using more efficient cryptography and data transport techniques which suits for an environment where fundamental limitations over memory, processor exists. . The proposed model can be utilized not only in M-commerce but also diverse wireless data communication such as mobile hospital and government, based on mobile phone through wireless internet.